# Retinal Malady Classification using AI:
# A novel ViT-SVM combination architecture.

Shashwat Jha
Department of Electrical and Electronics Engineering
Birla Institute of Technology Mesra
shashwat6jha@gmail.com

Vishvaditya Luhach
Department of Computer Science and Engineering
Maharaja Surajmal Institute of Technology
vishvadityaluhach@gmail.com

Dr.Raju Poddar
Department of Bio-Technology
Birla Institute of Technology Mesra
rpoddar@bitmesra.ac.in

*Abstract*— Macular Holes, Central serous retinopathy and Diabetic Retinopathy are one of the most widespread maladies of the eyes responsible for either partial or complete vision loss, thus making it clear that early detection of the mentioned defects is detrimental for the well-being of the patient. This study intends to introduce the application of Vision Transformer and Support Vector Machine based hybrid architecture (ViT-SVM) and analyse its performance to classify the optical coherence topography (OCT) Scans with the intention to automate the early detection of these retinal defects.

*Keywords*—ViT-SVM, Artificial intelligence (AI), Vision Transformer (ViT), Support Vector Machine (SVM), Hybrid, OCT -B, Retinal defect classification, Deep learning, Machine learning.

## 1. INTRODUCTION

Diabetic retinopathy is a microvascular disease of the retina and an terminal-organ problem of diabetes [1]. An important cause of blindness, it affects three out of four diabetic patients after 15 years of disease duration [2]. Central serous chorioretinopathy is a disease caused by the detachment of serous in the neurosensory retina. Its frequency is approximately 10 cases per 100,000 and is sixfold higher in men than in women. choroidal vascular hyperpermeability is considered as one of the main reasons for this [3]. Macular hole is an opening in the foveal centre of the retina. In most cases it is due to abnormal Vitreofoveal traction. Macular holes generally occur in the sixth to seventh decade of life of a patient [4].

Considering the prevalence of these diseases, the development of swift and accurate classification techniques for detection and classification are of utmost importance.

## 2. LITERATURE REVIEW

Since a long time, the classification tasks in cases of medical diagnosis of the eye and have been done manually by doctors, but recently a lot of work has been done on the attempt to automate this task with the assistance of artificial intelligence techniques for instance, Raja et al. proposed a CNN based Transfer Learning approach along with graph search to diagnose glaucoma [5]. whereas Mishra. et al presented Multi-Level Dual-Attention Based CNN for classification of Macular abnormalities using OCT scans [6]. Applications of Transfer Learning with CNN based models have also been discussed by Qomariah. et al for determining diabetic retinopathy with the additional assistance of SVM [7]. In fact, Huang. et al proposed a complete end to end solution for AMD diagnosis and treatment planning, based on Transfer Learning techniques [8]. A lot of work has been also presented on AI based retinal disease classification using coloured fundus images, Xu. et al [9] as well as Li. et al [10] have applied deep Convolutional as well as Multi-Instance Learning techniques for the diagnosis and classification.

Transformers which were first introduced in 2017 by Vaswani. et al have been nothing less than a breakthrough in attention based models being the first transduction model relying entirely on self-attention for computation of its input and output without using sequence aligned RNNs or Convolutions [11] with remarkable performance in the fields of natural language processing as well as time series based problems they have also proven their mettle in terms of their applications on images [12]. The Vision Transformer was initially proposed in 2021 by Dosovitskiy. et al which achieved an accuracy of 90.45% on the ImageNet Dataset and is still the second-best image classification architecture in the world [13].

Support Vector Machine (SVM), first introduced in 1995 [14] is one of the most widely utilised and celebrated techniques in history and is especially popular in classification problems. In terms of its applications in classifying retinal diseases it has been used in multiple instances such as Yang et al [15] in his research presented a hybrid Machine Learning model for automatic classification of retinal diseases U-net, PCA and SVM in combination. Anantrasirichai et al [16] utilised SVM based techniques for texture classification and Arunkumar et al [17] proposed SVM to classify the retinal diseases by reduced deep learning features.

Previously Vision Transformer (ViT) has only been utilised sporadically in biomedical applications. Wang et. al proposed the implementation of Vision Transformer for diagnosis of Genitourinary Syndrome of Menopause [18], whereas Wu et. al implemented this architecture for grade recognition of diabetic retinopathy [19].

Considering all the prior findings it can be argued that convolution based deep learning approaches and Machine learning algorithms have performed well on high computational cost and larger sized models on OCT -B scans or compromised on performance while reducing size and computational cost. Classification on OCT-B scan itself is an arduous task as the colour scale and intensity of the data makes it difficult for models to perform well. With our hybrid approach we intend to utilise the exceptional efficiency of SVM in terms of memory as well as dealing with datasets with higher dimensions and the noteworthy performance of attention based Vision transformer which still stands as one of the best image classification algorithms globally.

With this paper we intend to classify our OCT-B scan dataset into 4 classes. Normal, Macular Holes, Central serous retinopathy and Diabetic Retinopathy by implementing our Vision Transformer based proposed architecture (ViT-SVM) and then further comparing its performance on our dataset with only an ViT model with dense classification layer and VGG-16 pre-trained model.

## 3. DATASET AND PRE-PROCESSING

### 3.1. DATASET

The Dataset utilised for our research purposes, have been sourced in 4 classes of OCT – B scans with a total of 517 images of size 750x500. The dataset was created by Lakshminarayanan et. al which are publicly available at University of Waterloo data verse [20][21][22][23]. Figure.1 depicts samples of OCT-B scans including normal as well as the other disease-based classes of the scans.

The image data was loaded into a pandas dataframe and encoded into 4 classes as described below:

0- Retinal scans with central serous retinopathy consisting of 102 images.
1- Retinal scans with Diabetic Retinopathy consisting of 107 images.
2- Scans of retina with macular holes consisting of 102 images.
3- OCT scans of normal Retina consisting of 206 images.

### 3.2. PRE-PROCESSING

Before analysing the performance of deep learning models on the dataset, the images were first pre-processed. Images were pre-processed initially by resizing all images to 256 x 256 then by defining a custom pre-processing function and passing it into the ImageDataGenerator function of the KerasAPI in batches with batch size of 8 images each, which flipped the images along the vertical and the horizontal axis randomly.

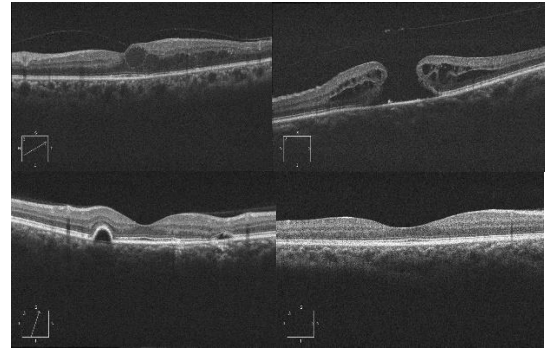

**Figure.1:** OCT-B scan samples from dataset depicting all four classes of images.

## 4. METHODOLOGY AND IMPLEMENTATION

The experimentation has been implemented using Tensorflow-gpu v2 [24] while using the Keras API [25]. All models were trained for 50 epochs and then performance was analyzed for the classification and diagnosis of Retinal diseases. The hardware specifications of the system on which the models were trained comprised of AMD Ryzen 5 3600 CPU, 16 GB of Ram and Nvidia 1660ti GPU was utilized for acceleration.

### 4.1. Vision Transformer (ViT)

The ViT is a visual Transformer architecture which represents an input image as a series of image patches and further predicts class labels for the image.

It is different from classical CNNs as they use pixel arrays for computations. On the contrary, Vision Transformer splits an image into fixed-size patches, further it inputs the patches into the Embedding layers of Linear Projections of Flattened Patches thereby generating vectors, usually called tokens. These tokens are added in front of a series of tokens. In addition, the location information needs to be added, corresponding to the position embeddings as in depicted in Figure.2. Then these tokens along with the location information is forwarded into the Transformer Encoder as embedded patches. Transformer Encoder has as many outputs as there are inputs. Finally, only classification is performed, so the output corresponding to the class position is input into MLP Head for prediction and classification output as depicted in Figure.2.

The Transformer encoder comprises of:
*Multi-Head Self Attention Layer (MSP):* This layer is responsible for the integration of the attention outputs linearly. The first thing that happens is a sequence of embeddings is passed into the encoder, the embeddings experience three separate linear transformations resulting in three vectors namely 'query', 'key' and 'value'. Now these 3 vectors are utilised to calculate the attention outputs for each embedding by taking dot product of the 3 vectors. Self-attention is computed multiple times in parallel and independently. It is therefore referred to as Multi-head Attention. The attention measures the strength of the relationship between the patches and then helps in the prediction. Figure.4 depicts the functioning of the MSP layer.

*Multi-Layer Perceptron (MLP) Layer:* This layer includes two layers with Gaussian Error Linear Unit (GELU).
*Layer Norm (LN):* This is included before each block, since it does not have any previous dependencies of images it improves the performance and reduces the time taken.
The construction and internal architecture of the Transformer encoder is depicted in Figure.3

The Vision Transformer model used was ViT-B32 which stands for Vision Transformer Base 32, it has patch resolution of 32 x 32. with 'Softmax' activation. Learning rate was set as 1e-4 initially and was reduced during training. Further, a small neural network was attached through a 'Flatten' layer which was added to flatten the output of the Vision Transformer into a 1-D array, a 'Dropout' layer with value 0.5 was connected and at last a dense layer with 4 neurons and 'Softmax' activation function and 'Adam' optimizer was added for classification.

*4.2 TRANSFER LEARNING - (VGG-16)*

The performance of our proposed Vision Transformer model was compared with comparable sized popular classification algorithm VGG-16. VGG-16 is classical Convolutional neural network architecture. The network utilizes small 3 x 3 filters. Otherwise considered a simple architecture the other components being pooling layers and a fully connected layer being 16 layers deep [26].

The VGG-16 model have been applied using pre-trained weights from ImageNet dataset [27]. The input shape for VGG-16 model was same as the Vision Transformer, i.e. 256x256x3. The output of VGG-16 model was further fed into a 'Flatten' layer, then a Dropout layer with a dropout rate of 0.5 and at last a Dense layer with 'softmax' activation was used for classification. Both the models were compiled with categorical cross-entropy as the loss function and using Adam optimizer.

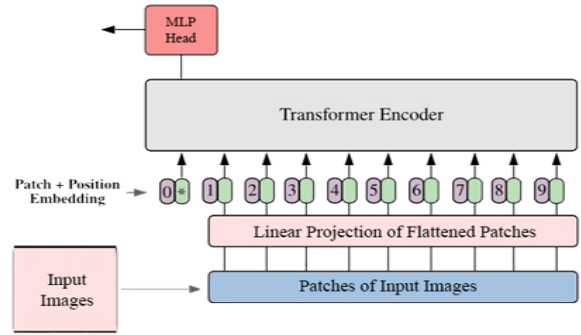

**Figure.2**: Architecture of Vision Transformer

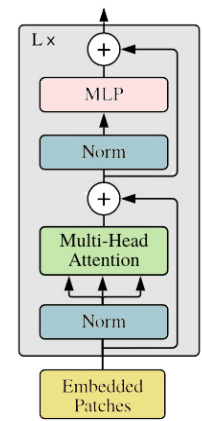

**Figure.3:** Internal architecture of Transformer Encoder

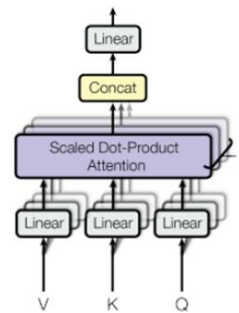

**Figure.4:** The Multi-Head Self Attention Layer

*4.3 PROPOSED ViT-SVM ARCHITECTURE*

The architecture proposed in this paper presents a robust hybrid architecture (ViT-SVM) combining the attention based approach of the transformers with the memory efficient, decision function based SVM.

The first block of the proposed model consists of the Vision Transformer. The vision transformer block utilised here was the ViT-B32 architecture, with a patch resolution of 32x32 and consisting the same sequence of porcedure as mentioned in the section (4.1) till the MLP head with 'Softmax' activation function. The output of the Vision transformer was fed into the SVM by connecting 2 Dense layers, one with 64 Neurons and the other having 4 neurons for classification. L2 Regularization was performed with a regularization factor of 0.01, 'Squared hinge' was defined as the loss function along with 'Adam' optimizer. The final dense layer had a 'softmax' activation function for the multi-class classification. Figure.5 represents the architecture of the proposed ViT- SVM model.

## 5. RESULTS

All the models were analyzed on various performance metrics namely, precision, recall and accuracy presented decent performance on this classification problem.VGG-16 after training achieved an accuracy of 83% on the test set, On the other hand Vision Transformer achieved an accuracy of 88%. Our proposed ViT-SVM architecture outperformed all other models in the experiment achieving an accuracy of 94% in classifying the mentioned Retinal maladies. In terms of Precision which is the ratio of actual true cases and the total cases predicted true by the model as well as Recall which is the ratio of true positive cases and the total actual positive cases, VGG-16 achieves perfect precision in predicting normal as well as retina with diabetic retinopathy while achieving a recall of 0.90 in predicting central serous retinopathy and Macular hole. ViT on the other hand achieves a precision score of 0.90 in detecting diabetic retinopathy and central serous retinopathy while also achieving a perfect recall score for normal retina and 0.90 recall in detecting central serous retinopathy. Our proposed hybrid architecture achieves a perfect precision and recall score in diagnosing normal retina and diabetic retinopathy and a recall of 0.90 in detecting macular hole.

Figures.6-8, depict the precision and recall of each model in the graphical format and Figure.9 represents the cross-model comparison of each model in terms of accuracy. Table.1 consists of the complete comparison in terms of all the metrics in the tabular format whereas Figure 10 represents the confusion matrix of the proposed hybrid ViT-SVM model depicting true values and predicted values for each class.

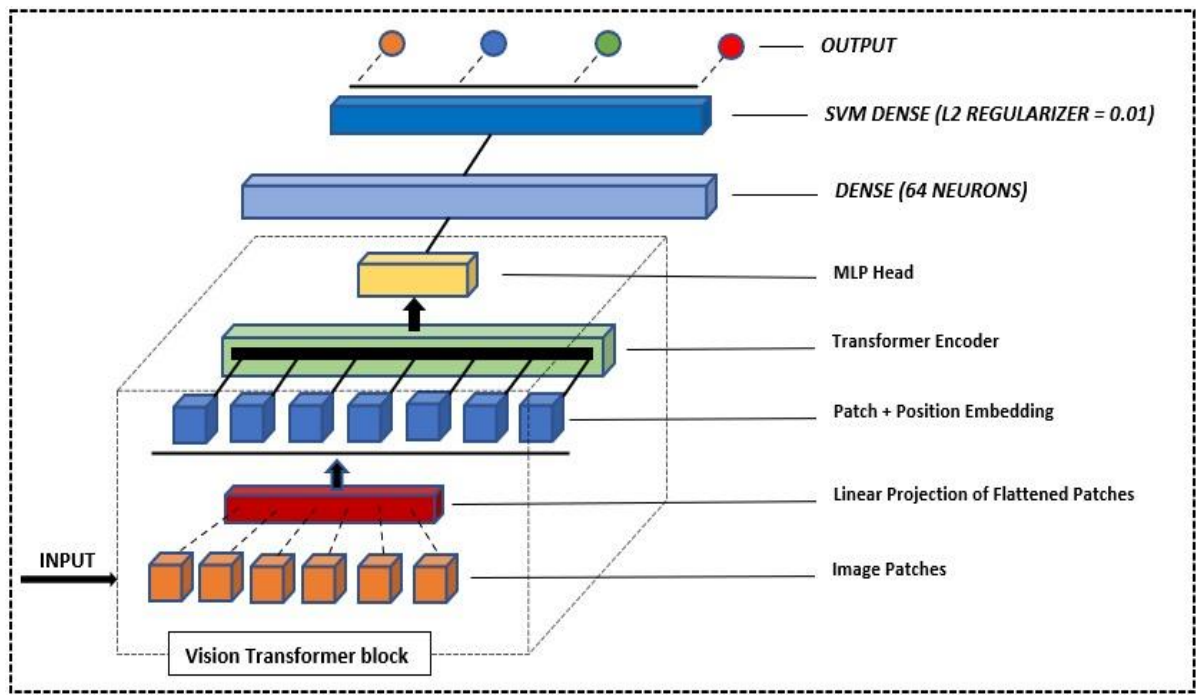

**Figure 5**: Architecture of Proposed Hybrid (ViT-SVM) Model.

**Table 1**: Result Comparison of the Models

| Model | Normal (No Maladies) | | Diabetic Retinopathy | | Central-Serous Retinopathy | | Macular Hole | | Accuracy |
|---|---|---|---|---|---|---|---|---|---|
| | Precision | Recall | Precision | Recall | Precision | Recall | Precision | Recall | |
| VGG-16 | **1.00** | 0.95 | **1.00** | 0.45 | 0.64 | **0.90** | 0.69 | **0.90** | 83% |
| ViT | 0.88 | **1.00** | 0.90 | 0.82 | **0.90** | **0.90** | 0.88 | 0.70 | 88% |
| ViT+SVM Hybrid | **1.00** | **1.00** | **1.00** | **1.00** | 0.89 | 0.80 | 0.82 | **0.90** | **94%** |

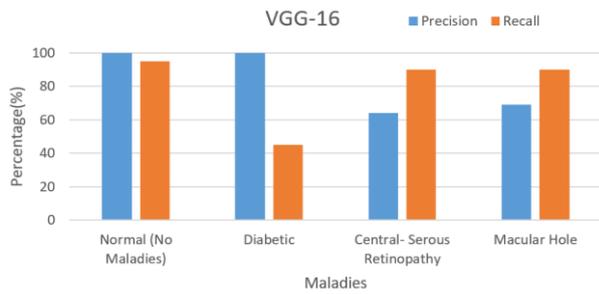

**Figure.6:** Precision and Recall representation for VGG-16

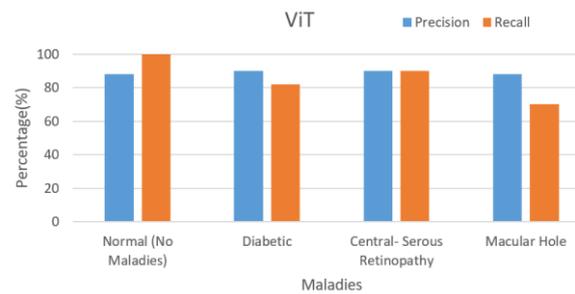

**Figure.7:** Precision and Recall representation for ViT

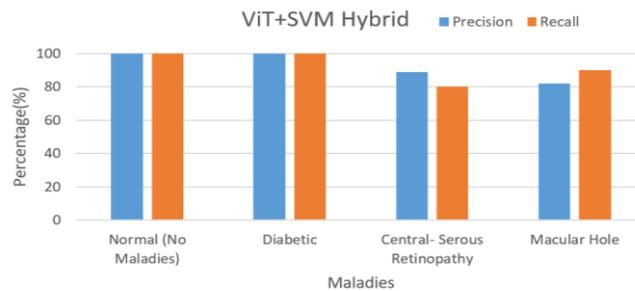

**Figure.8:** Precision and Recall representation for ViT-SVM

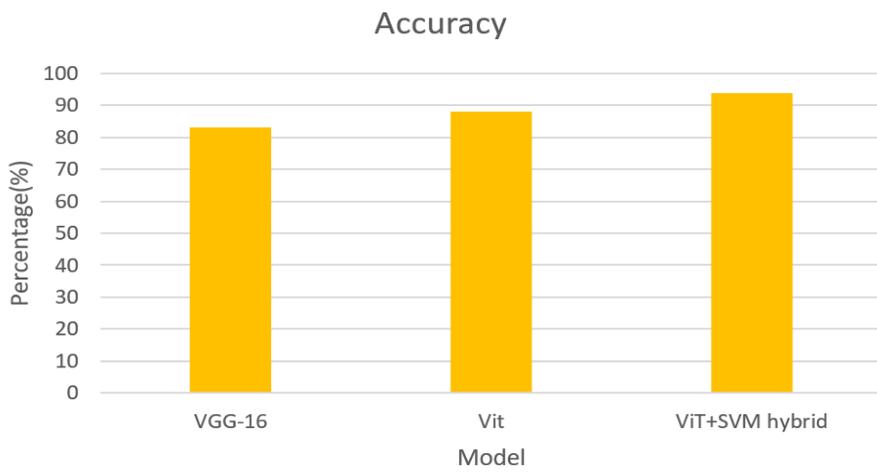

**Figure.9:** Cross model performance - Accuracy

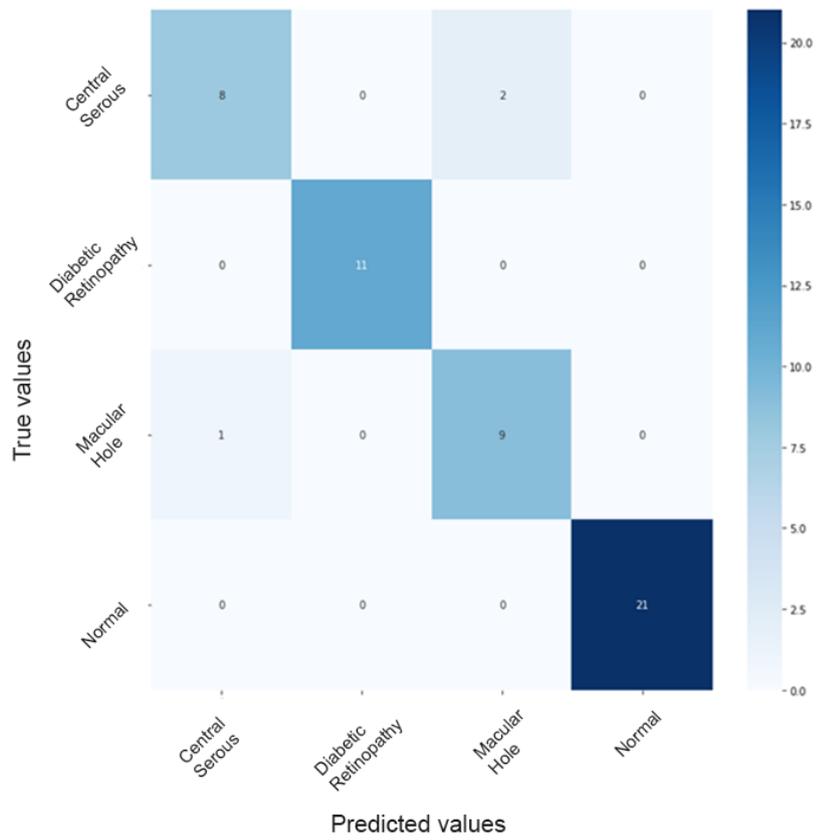

**Figure.10:** Confusion Matrix for ViT-SVM

## 6. CONCLUSION

After analysis of models, it is certain that Transformer based have proven to be effective in diagnosing as well as classifying the mentioned retinal diseases and have performed decently considering the size and the dimensions of the dataset. The proposed ViT-SVM model had the best result followed by the performance of the ViT model. VGG-16 performed the poorest in this experiment. Considering the size of each model as well as the computational cost, one can argue that the Vision Transformer based architectures are of greater caliber and are more efficient than classical Convolutional based VGG-16. Moreover, introduction of Support Vector Machine along with the Vision transformer had performed best by a margin of 6% over the simple ViT and 11% over VGG-16 in terms of overall accuracy.

Since it is evident that ViT based models incorporate great caliber within themselves, future research is encouraged as one can fine tune these architectures further as well as improve the quality and size of dataset for better performance and analysis.

## 7. REFERECNES